# Coulomb blockade in field electron emission from carbon nanotubes

Victor I. Kleshch[1,*], Vitali Porshyn[2], Pavel Serbun[2], Anton S. Orekhov[1,3,4], Rinat R. Ismagilov[1], Sergey A. Malykhin[1,5,6], Valentina A. Eremina[4,7], Elena D. Obraztsova[4,7], Dirk Lützenkirchen-Hecht[2], and Alexander N. Obraztsov[1,5]

[1]*Department of Physics, Lomonosov Moscow State University, Moscow 119991, Russia*

[2]*Physics Department, Faculty of Mathematics and Natural Sciences, University of Wuppertal, Wuppertal 42119, Germany*

[3]*NRC Kurchatov Institute, 123182 Moscow, Russia*

[4]*Moscow Institute of Physics and Technology, 141701 Dolgoprudny, Moscow Region, Russia*

[5]*Department of Physics and Mathematics, University of Eastern Finland, 80101 Joensuu, Finland*

[6]*Department of Solid State Physics, Lebedev Physical Institute, RAS, 119991 Moscow, Russia*

[7]*Prokhorov General Physics Institute, RAS, 119991 Moscow, Russia*

* Corresponding author, e-mail: klesch@polly.phys.msu.ru

We report the observation of Coulomb blockade in field electron emission (FE) from single-wall carbon nanotubes (SWCNTs), which is manifested as pronounced steps in the FE current-voltage curves and oscillatory variations in the energy distribution of emitted electrons. The appearance of the Coulomb blockade is explained by the formation of nanoscale protrusions at the apexes of SWCNTs due to the electric field-assisted surface diffusion of adsorbates and carbon adatoms. The proposed adsorbate-assisted FE mechanism is substantially different from the well-known resonant tunneling associated with discrete electronic states of adsorbed atoms. The simulations based on the Coulomb blockade theory are in excellent agreement with the experimental results. The SWCNT field emitters controlled by the Coulomb blockade effect are expected to be used to develop on-demand coherent single-electron sources for advanced vacuum nanoelectronic devices.

**1. Introduction**

The highest brightness, coherence and monochromaticity of field emission (FE) electron sources make them important for various fundamental studies and applications. Recent advances in this area include significant improvements in the performance of conventional FE cathodes based on metal nanotips [1], as well as the development of new types of field emitters based on nanoscale structures [2,3] and individual molecules [4]. Carbon nanotubes (CNTs) are considered as one of the most promising nanomaterials for the creation of FE cathodes [5,6] and are successfully used in commercial applications, for example, in portable x-ray tubes [7].

FE from CNTs has been extensively studied since 1995 [8]. It is well-established [9,10] that the main FE characteristics of CNT field emitters can be adequately described by the Fowler-Nordheim (FN) theory [11,12], which predicts linear FE current-voltage, $I(V)$, curves in FN coordinates (FN plots) and a single asymmetric peak in the energy spectrum of the emitted electrons. However, the experimental observation of such characteristics usually requires special treatment, such as annealing of CNTs at high temperatures of about 1000°C [13]. Otherwise, the effects related to the presence of adsorbates or carbonaceous impurities lead to strong deviations from the FN theory, including highly nonlinear FN plots with numerous kinks and multiple peaks in the electron energy spectra [14-17]. These deviations are usually explained in terms of a model based on resonant tunneling [18,19] through a double-barrier structure formed by adsorbates on the CNT apex [20]. However, an alternative explanation based on the Coulomb blockade and other single-electron charging effects [21,22], which can also affect the electron transport in double-barrier structures, were generally overlooked. The theory of FE in the Coulomb blockade regime was proposed by Raichev [23] more than a decade since the first observations of FE from CNTs. Experimental evidence for the Coulomb blockade of FE was recently demonstrated for ultrashort single-wall CNTs (SWCNTs) [24], which showed regular steps in the $I(V)$ curve (Coulomb staircase). Still, a similar staircase behavior was observed for other CNT field emitters [13,16] and was explained by resonant tunneling processes. Thus, it is difficult to establish the actual FE mechanism solely on the basis of *I*(*V*) measurements.

Here, we measure the energy distribution of the emitted electrons to probe electronic states in SWCNT field emitters, which exhibit well-defined steps in the *I*(*V*) curves. The energy spectra show that the Fermi level of the emitter oscillates with increasing voltage, indicating unambiguously that the FE is governed by the Coulomb blockade. We propose a model based on the assumption of FE-induced formation of nanoscale protrusions on the SWCNTs tips and perform numerical simulations, which show excellent agreement with the experiment. Finally, we compare our results with other studies of Coulomb-blockade-controlled field emitters and discuss possible applications of single-electron charging effects in vacuum nanoelectronic devices.

2. Methods

2.1. SWCNT samples

SWCNTs powders supplied by OCSiAl company [25] were used to fabricate thin films by means of the techniques described in detail in Ref. [26]. In brief, purification of the powder from byproducts followed by the separation of metallic and semiconducting nanotubes was performed. A preliminary investigation of thin films made from separated over conductivity type SWCNTs [26] showed that the specific features in the FE characteristics, studied in this work, are observed for both metallic and semiconducting SWCNTs. Therefore, in the following we present results obtained for samples of both types and do not specify the type of conductivity. Figure 1 shows typical scanning electron microscopy (SEM) images of a SWCNT film with an average thickness of 300 nm. To investigate the FE properties, the film was either transferred to a flat silicon substrate (Fig. 1a), or was fixed on a massive sharpened tungsten wire (Fig. 1b) by creating a platinum contact using focused ion-beam deposition. The edges of the films usually contained a number of protruding cone-shaped nanotube bundles with a length of about 500 nm (Fig. 1c, d).

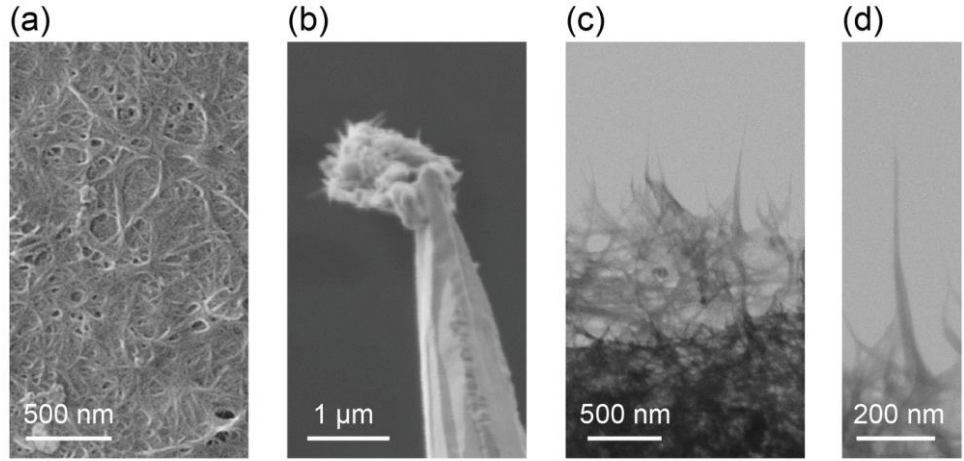

**Fig. 1.** Typical SEM images of a SWCNT film. (a) Top view of the film transferred to a flat silicon substrate. (b) A freestanding piece of the film fixed on a tungsten tip. (c) The edge of the freestanding film. (d) A cone-shaped nanotube bundle at the edge of the freestanding film.

**2.2. FE measurements**

In order to exclude possible artifacts and confirm the reproducibility of the FE measurements, we used two different ultrahigh vacuum (UHV) experimental setups: the scanning field emission microscopy (SFEM) setup and the field emission spectroscopy (FES) setup, which are described in Refs. [27,28]. In the SFEM setup, a tungsten needle anode was positioned at a distance of 20 μm above the surface of the SWCNT film transferred to a flat silicon substrate (Fig. 2a). In the FES setup, a piece of freestanding SWCNT film fixed on a tungsten holder was placed at a distance of 500 μm from the metal mesh anode (Fig. 2b). In both setups the DC voltage, *V*, was applied between the SWCNT film (cathode) and the anode, and the FE current from the cathode, *I*, was measured by picoammeter. Moreover, in the FES setup, the total electron energy distribution, $J(\varepsilon)$, was measured by an electron energy spectrometer located behind the mesh anode. The measurements were performed at a pressure of about $5\times10^{-8}$ Pa and at room temperature.

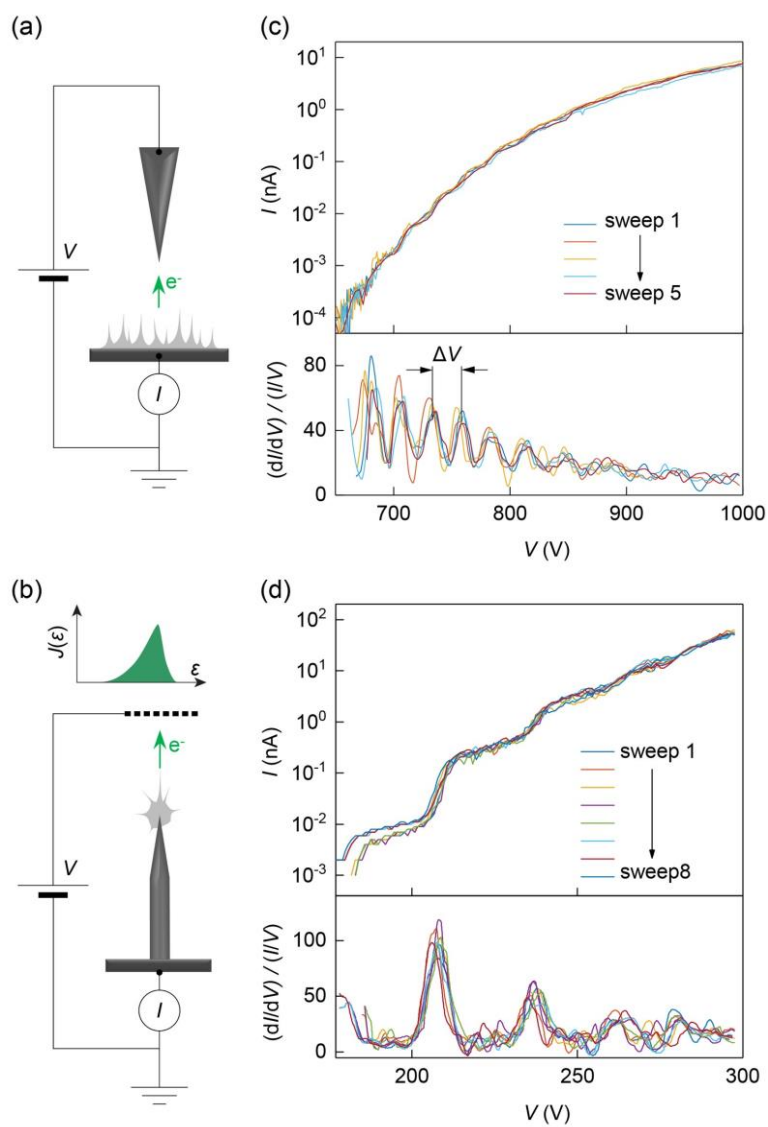

**Fig. 2.** Schematic diagrams of the SFEM (a) and FES (b) experimental setups. (c) The FE current-voltage characteristics with pronounced steps and the corresponding normalized differential conductivity curves obtained in the SFEM setup for the SWCNT film shown in Fig. 1a. (d) The same type of FE data as in (c) obtained in the FES setup for the SWCNT sample presented in Fig. 1b.

## 3. Results

Initially, the FE current from a SWCNT film was unstable due to the presence of adsorbates on the surface. In order to stabilize the emission, a series of voltage cycles (conditioning) was performed (See Supplementary Fig. S1). Often, as a result of the conditioning process, the FE $I(V)$ curves exhibited regular steps, which were well reproducible at currents not exceeding 100 nA. Typical *I*(*V*) curves obtained in the SFEM and FES setups are presented in Fig. 2c and Fig. 2d respectively. The corresponding curves of normalized differential conductivity, (d*I*/d*V*)/(*I*/*V*), show oscillations with a period $\Delta V$, varying in the range from 10 to 100 V for different regions on the surface of the SWCNT film. The observed behavior of the FE current during conditioning and the parameters of steps in the *I*(*V*) curves were qualitatively the same for both experimental setups. Furthermore, the voltage values used in the FES setup were noticeably lower, due to the additional field enhancement provided by the needle-like tungsten sample holder.

Figure 3a shows the voltage dependence of the energy spectra corresponding to the *I*(*V*) curve presented in Fig. 4a. Each spectrum is normalized to unity, and the electron kinetic energy, $\varepsilon$, is measured relative to the Fermi level of the cathode. One can see that at $V = 250$ V a single peak in the energy spectrum is observed at $\varepsilon \approx 0$ eV. This peak shifts with increasing voltage, and when the position of the peak reaches $\varepsilon \approx -0.5$ eV, its intensity abruptly decreases. At the same time, a new peak develops at $\varepsilon \approx 0$ eV, which continues to shift with a further increase in voltage, and the process repeats again. There is a clear correlation between the dependence of the peak position (Fig. 4c) and the corresponding normalized differential conductance (Fig. 4b). Each maximum in the periodically oscillating differential conductance oscillations is reached when a new peak develops in the $J(\varepsilon)$ spectrum. It should be noted, that the area under the spectrum before normalization (integrated count rate) is proportional to the total current measured by the picoammeter (see Supplementary Fig. S2). This linear correlation between the FE current and the count rate strongly suggests that the FE solely occurred from a single point emitter (i.e. a single nanotube), which contributed to the total current and was measured by the spectrometer.

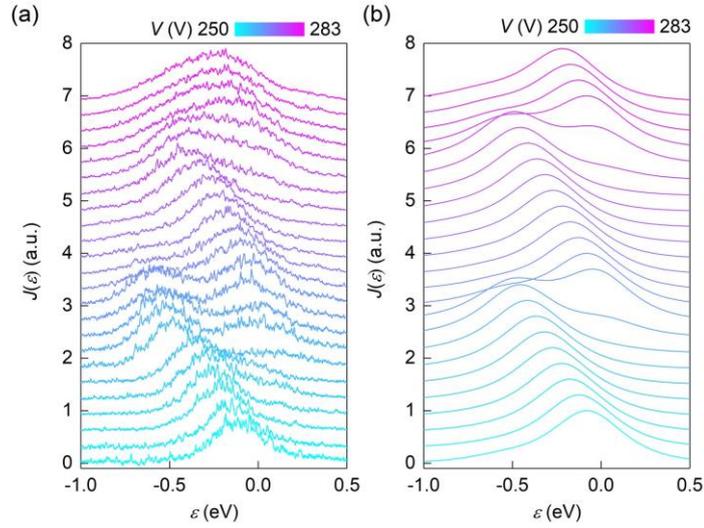

**Fig. 3.** (a) Experimental dependence of the normalized total electron energy distributions on the applied voltage measured for the SWCNT sample shown in Fig. 1b. (b) The corresponding simulated energy distributions.

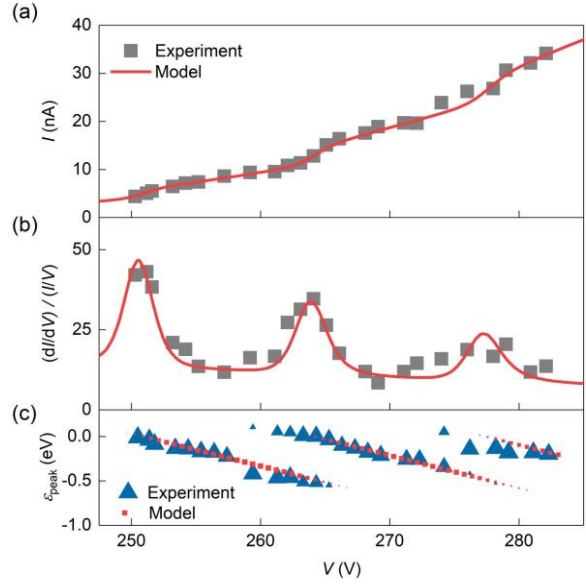

**Fig. 4.** (a) Current-voltage characteristic and (b) the corresponding normalized differential conductivity obtained simultaneously with the spectra, $J(\varepsilon)$, shown in Fig. 3a. (c) Voltage dependence of the energy peak positions, $\varepsilon_{peak}(V)$, in Fig. 3a, determined from the $J(\varepsilon)$ curve fitting (see Supplementary Fig. S3). The size of each point in $\varepsilon_{peak}(V)$ is proportional to the amplitude of the corresponding energy peak. The red lines in (a, b) and red squares in (c) are fits of the experimental dependencies obtained using the model of FE in the Coulomb blockade regime. See text for more details.

## 4. Discussion

The oscillatory behavior of the energy peak positions with voltage, observed in our experiments, is quite unusual for CNT-based field emitters. More often, energy spectra show a continuous linear shift with the applied voltage of one or several energy peaks [15,29] rather than oscillations. This shift is commonly explained by an electric field-induced change in energy levels, corresponding to the localized electronic states in the CNT cap [30] or states associated with adsorbed atoms and molecules [31]. Recently, we have shown that the energy peak oscillations is a clear fingerprint of the Coulomb blockade [32], which arises when FE occurs from a nanoscale structure separated by a tunnel barrier from the rest of the cathode [23]. The transfer of an electron through the tunnel barrier leads to a discrete change in the electric field on the surface of the nanostructure and the corresponding abrupt changes in the FE current and energy spectrum. Thus, we assume that in the measurements shown in Figs. 2-4 FE occurred from isolated nanostructures located at the tips of the SWCNTs. Such nanostructures, usually called nanoprotrusions, can self-assemble at the apex of a nanotube during the conditioning process, due to an electric field-assisted surface diffusion [33] of carbonaceous impurities or adsorbed atoms and molecules. The difference in the electronic properties of well-conducting SWCNT and poorly conducting amorphous nanoprotrusion leads to the formation of a potential barrier between them. In addition, the electrical isolation of nanoprotrusions can be associated with the presence of residual polymers and surfactants (used for purification of nanotube powders [26]) on the surface of SWCNTs.

We performed numerical simulations of the FE characteristics of a SWCNT terminated with a nanoprotrusion, using the model of FE in the Coulomb blockade regime, which is presented in detail in Refs. [23,32]. In this model, the electron transport is described by the master equation [21] that gives the probabilities $P(N)$ for the nanoprotrusion to be in the state with $N$ electrons at a fixed applied voltage. The total energy spectrum and the FE current are then calculated as the sum of the partial currents and spectra corresponding to each $N$-state, multiplied by $P(N)$. The partial currents and spectra are determined by the electric field distribution, which was calculated by numerically solving the Laplace equation, using the actual dimensions of the SWCNT film and the electrodes used in the experimental setup.

The geometry of the model and the results of the electrostatic simulations are presented in Fig. 5. The cathode consists of a sharpened tungsten holder (Fig. 5a) with a cone-shaped structure at the top (Fig. 5b), representing a SWCNT bundle, similar to that shown in Fig. 1d. The bundle ends with a nanotube having a hemispherically capped cylindrical nanoprotrusion on its apex (Fig. 5c). The nanoprotrusion with a height $H$ is separated from the nanotube by an insulating layer with a thickness $h$, representing the tunnel barrier region. The nanotube diameter, $D$, is of 1.8 nm, which is an average diameter value for this type of SWCNTs [26]. The dielectric constant of the insulating layer was chosen to be 5, corresponding to the typical value for amorphous carbon (a-C) [34]. The parameters $H$ and $h$ were varied during the modeling in order to achieve agreement between the capacitive characteristics of the system, determined from the experimental data and from the field distribution simulations. The capacitance of the nanoprotrusion with respect to the anode, $C_a$, is related to the period of the differential conductance oscillations as $\Delta V = e/C_a$, where $e$ is the elementary charge. The total capacitance of the nanoprotrusion, $C$, is related to the charging energy $E_c = e^2/C$ [22], i.e. the energy cost for an electron transfer to the nanoprotrusion. The $E_c$ value is determined by the amplitude of the $\varepsilon_{peak}(V)$ dependence. From the experimental results presented in Fig. 4, it follows that $E_c = 0.5$ eV and $\Delta V = 13$ V, and, therefore, $C = 320$ zF and $C_a = 12.3$ zF. The best match with these experimental values of $C$ and $C_a$ was obtained by the simulations at $H = 3.3$ nm and $h = 0.7$ nm.

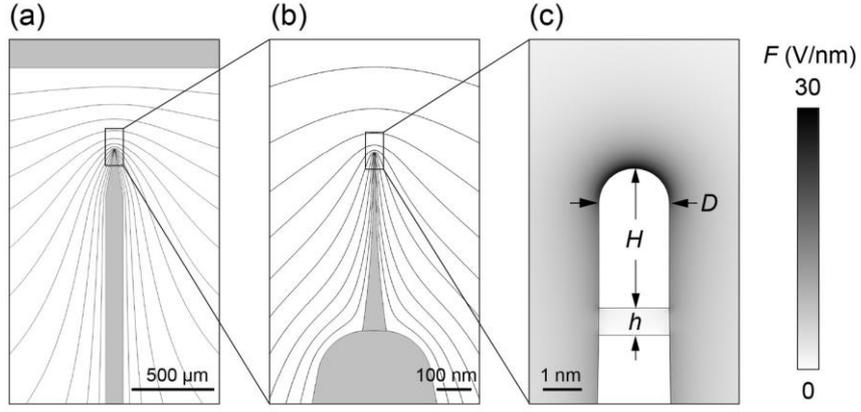

**Fig. 5.** Electrostatic modeling for a SWCNT-based emitter at an anode voltage $V=260$ V. (a) Cross-section of the space between the tip cathode and the flat anode with simulated equipotential lines. The potential difference between the neighboring lines is $\Delta U=20$ V. (b) Zoom-in to the apex region, showing the cone-shaped SWCNT bundle and equipotential lines with $\Delta U=5$ V. (c) Distribution of the electric field strength in the vicinity of a hemispherically capped cylindrical nanoprotrusion on top of a SWCNT.

The simulated energy spectra (Fig. 3b) and the FE currents (Fig. 4a) reproduce well the experimentally observed behavior. In particular, an abrupt suppression of the peak at $\varepsilon = -E_c$ is observed, when a new energy peak appears at $\varepsilon = 0$ eV. It is important to note, that in the case of FE via the resonant tunneling process, additional peaks can also appear with increasing voltage, when the energy level of a resonant state becomes aligned with the Fermi level in the nanotube. However, there is no suppression of the underlying energy peaks, instead, all the resonant peaks continue to shift downward in energy as the voltage is increased further [15,35]. In the case of FE in the Coulomb blockade regime the mechanism is substantially different: the FE from the state, corresponding to the peak at $\varepsilon = -E_c$, becomes suppressed, since the probability, $P_0$, of transition to the state at $\varepsilon = 0$ eV by the transfer of an electron through the internal barrier is much higher than the probability, $P_{FE}$, of electrons tunneling to the vacuum. The probabilities $P_{FE}$ and $P_0$ are proportional to the corresponding tunneling rates, $\Gamma_{FE}$ and $\Gamma_0$, which can be roughly estimated as $\Gamma_{FE} \approx I/e$ and $\Gamma_0 \approx (RC)^{-1}$ [32], where $R$ is the tunneling resistance of the internal barrier. $R$ is the fitting parameter in the modeling and was determined to be of the order of 1 MΩ. Thus, using the $C$ value determined above and the typical FE current value of $I$=10nA, one has $P_0 \sim \Gamma_0 \approx 3 \times 10^{12} \text{s}^{-1}$, which is much higher than $P_{\text{FE}} \sim \Gamma_{\text{FE}} \approx 0.06 \times 10^{12} \text{s}^{-1}$, meaning that the condition for the suppression of the energy peak, $P_0 \gg P_{\text{FE}}$, is well met.

Let us now compare our results with previous studies on the Coulomb blockade effect in FE. In the pioneering work by Pascale-Hamri et al. [24], the Coulomb staircase in $I(V)$ curves was observed for ultrashort (less than 60 nm long) SWCNTs grown on a tungsten tip covered by an insulating a-C layer. More recently Kleshch et al. [32] reported electron energy spectroscopy measurements of Coulomb blockade of FE from carbon nanowires formed on the surface of an a-C layer covering a diamond nanotip. All the parameters $E_c$, $\Delta V$, $C$, $C_a$, etc. obtained in these studies are in the same range as in the present experiments, showing that the mechanism of FE very likely is the same. However, both ultrashort SWCNTs and carbon nanowires showed much higher FE currents up to several microamperes and superior stability, which is a consequence of their robust structure, provided by covalent carbon-carbon bonding. In contrast, nanoprotrusions are more fragile and break at current levels of about 100 nA. Nevertheless, during conditioning, nanoprotrusions become quite stable below 100 nA, which allows to perform detailed FE measurements and establish the electron transport mechanism. Finally, it is important to note that the formation of nanoprotrusions and the oscillatory behavior of energy peaks, similar to that observed here, have previously been reported in the literature, for example in the work by Purcell et al. [13] on FE from multi-wall CNTs. This indicates that the single-electron charging effects resulting from nanoprotrusions formation are not specific to our SWCNT films, but are a general phenomenon for CNT-based field emitters.

## 5. Conclusion

In conclusion, we have presented strong evidence that the formation of nanoprotrusions at the apexes of SWCNTs leads to the Coulomb blockade of FE, which causes current steps in the FE current-voltage curves due to single-electron charging of a nanoprotrusion. The charging energy was measured directly by the electron energy spectroscopy to be 0.5 eV, which significantly exceeds the thermal fluctuations energy at room temperature and makes it possible to observe single-electron tunneling effects without cooling. The numerical simulations reproduce well the experimental results and indicate that the characteristic size of the nanoprotrusions is on the order of few nanometers. An important feature of SWCNT field emitters controlled by the Coulomb blockade is that electrons are emitted one by one at a fixed applied voltage. This can be used to create coherent single-electron sources for advanced vacuum nanoelectronic devices. For example, ultrafast laser-induced gating [36] of a Coulomb-blockade-controlled field emitter can provide on-demand ultrashort electron bunches, which are of great interest for time-resolved electron microscopy [37] and electron diffraction [38] techniques.

**Acknowledgements**

V.I.K., A.S.O., R.R.I. and S.A.M. acknowledge the support of the field emission experiments, modeling and electron microscopy studies by the Russian Science Foundation (Grant No. 19-72-10067). Separation of nanotubes over conductivity type and film formation was supported by RSF grant 20-42-08004. V.P. and D.L.-H. are grateful for German Federal Ministry of Education and Research (Project No. 05K13PX2) for infrastructure support.

**Author Contributions**

E.D.O and V.A.E. fabricated the SWCNT films. V.I.K. performed the FE experiments, assisted by V.P. and P.S. A.S.O. performed SEM characterization. V.I.K. analyzed the data and performed simulations, assisted by R.R.I. and S.A.M. V.I.K., D.L.-H. and A.N.O. co-wrote the manuscript with input from all co-authors.

**References**

[1] S. Tsujino, P. Das Kanungo, M. Monshipouri, C. Lee, and R. J. D. Miller, *Measurement of transverse emittance and coherence of double-gate field emitter array cathodes*, Nat. Commun. **7**, 13976 (2016).

[2] H. Zhang, J. Tang, J. S. Yuan, Y. Yamauchi, T. T. Suzuki, N. Shinya, K. Nakajima, and L. C. Qin, *An ultrabright and monochromatic electron point source made of a LaB6 nanowire*, Nat. Nanotechnol. **11**, 273 (2016).

[3] X. Y. Shao, A. Srinivasan, W. K. Ang, and A. Khursheed, *A high-brightness large-diameter graphene coated point cathode field emission electron source*, Nat. Commun. **9**, 1288 (2018).

[4] T. Esat, N. Friedrich, F. S. Tautz, and R. Temirov, *A standing molecule as a single-electron field emitter*, Nature **558**, 573 (2018).

[5] J. M. Bonard, M. Croci, C. Klinke, R. Kurt, O. Noury, and N. Weiss, *Carbon nanotube films as electron field emitters*, Carbon **40**, 1715 (2002).

[6] C. M. Collins, R. J. Parmee, W. I. Milne, and M. T. Cole, *High Performance Field Emitters*, Adv. Sci. **3**, 1500318 (2016).

[7] C. Puett, C. Inscoe, A. Hartman, J. Calliste, D. K. Franceschi, J. P. Lu, O. Zhou, and Y. Z. Lee, *An update on carbon nanotube-enabled X-ray sources for biomedical imaging*, Wiley Interdiscip. Rev.: Nanomed. Nanobiotechnol. **10**, e1475 (2018).

[8] J. M. Bonard, H. Kind, T. Stockli, and L. A. Nilsson, *Field emission from carbon nanotubes: the first five years*, Solid·State Electron. **45**, 893 (2001).

[9] S. T. Purcell, P. Vincent, C. Journet, and V. T. Binh, *Hot nanotubes: Stable heating of individual multiwall carbon nanotubes to 2000 K induced by the field-emission current*, Phys. Rev. Lett. **88**, 105502 (2002).

[10]     J. M. Bonard, K. A. Dean, B. F. Coll, and C. Klinke, *Field emission of individual carbon nanotubes in the scanning electron microscope*, Phys. Rev. Lett. **89**, 197602 (2002).

[11]     R. H. Fowler and L. Nordheim, *Electron emission in intense electric fields*, Proc. R. Soc. London, Ser. A **119**, 173 (1928).

[12]     R. D. Young, *Theoretical total-energy distribution of field-emitted electrons*, Phys. Rev. **113**, 110 (1959).

[13]     S. T. Purcell, P. Vincent, M. Rodriguez, C. Journet, S. Vignoli, D. Guillot, and A. Ayari, *Evolution of the field-emission properties of individual multiwalled carbon nanotubes submitted to temperature and field treatments*, Chem. Vapor Depos. **12**, 331 (2006).

[14]     J. D. Jarvis, H. L. Andrews, B. Ivanov, C. L. Stewart, N. de Jonge, E. C. Heeres, W. P. Kang, Y. M. Wong, J. L. Davidson, and C. A. Brau, *Resonant tunneling and extreme brightness from diamond field emitters and carbon nanotubes*, J. Appl. Phys. **108**, 094322 (2010).

[15]     M. J. Fransen, T. L. van Rooy, and P. Kruit, *Field emission energy distributions from individual multiwalled carbon nanotubes*, Appl. Surf. Sci. **146**, 312 (1999).

[16]     S. M. Lyth and S. R. P. Silva, *Resonant behavior observed in electron field emission from acid functionalized multiwall carbon nanotubes*, Appl. Phys. Lett. **94**, 123102 (2009).

[17]     E. A. Vasil'eva, V. I. Kleshch, and A. N. Obraztsov, *Effect of vacuum level on field emission from nanographite films*, Tech. Phys. **57**, 1003 (2012).

[18]     L. L. Chang, L. Esaki, and R. Tsu, *Resonant tunneling in semiconductor double barriers*, Appl. Phys. Lett. **24**, 593 (1974).

[19]     B. Ricco and M. Y. Azbel, *Physics of resonant tunneling. The one-dimensional double-barrier case*, Phys. Rev. B **29**, 1970 (1984).

[20]     K. A. Dean, P. von Allmen, and B. R. Chalamala, *Three behavioral states observed in field emission from single-walled carbon nanotubes*, J. Vac. Sci. Technol. B **17**, 1959 (1999).

[21]     I. O. Kulik and R. I. Shekhter, *Kinetic phenomena and charge discreteness effects in granulated media*, Sov. Phys.-JETP. **41**, 308 (1975).

[22]   K. K. Likharev, *Single-electron devices and their applications*, Proc. IEEE **87** (1999).

[23]   O. E. Raichev, *Coulomb blockade of field emission from nanoscale conductors*, Phys. Rev. B **73**, 195328 (2006).

[24]   A. Pascale-Hamri, S. Perisanu, A. Derouet, C. Journet, P. Vincent, A. Ayari, and S. T. Purcell, *Ultrashort single-wall carbon nanotubes reveal field-emission Coulomb blockade and highest electron-source brightness*, Phys. Rev. Lett. **112**, 126805 (2014).

[25]   B. Krause, P. Potschke, E. Ilin, and M. Predtechenskiy, *Melt mixed SWCNT-polypropylene composites with very low electrical percolation*, Polymer **98**, 45 (2016).

[26]   V. I. Kleshch, V. A. Eremina, P. Serbun, A. S. Orekhov, D. Lutzenkirchen-Hecht, E. D. Obraztsova, and A. N. Obraztsov, *A comparative study of field emission from semiconducting and metallic single-walled carbon nanotube planar emitters*, Phys. Status Solidi B **255**, 1700268 (2018).

[27]   S. Mingels, V. Porshyn, B. Bornmann, D. Lützenkirchen-Hecht, and G. Müller, *Sensitive fast electron spectrometer in adjustable triode configuration with pulsed tunable laser for research on photo-induced field emission cathodes*, Rev. Sci. Instrum. **86**, 043307 (2015).

[28]   P. Serbun, V. Porshyn, G. Müller, and D. Lützenkirchen-Hecht, *Advanced field emission measurement techniques for research on modern cold cathode materials and their applications for transmission-type x-ray sources*, Rev. Sci. Instrum. **91**, 083906 (2020).

[29]   D. Lovall, M. Buss, E. Graugnard, R. P. Andres, and R. Reifenberger, *Electron emission and structural characterization of a rope of single-walled carbon nanotubes*, Phys. Rev. B **61**, 5683 (2000).

[30]   X. Zheng, G. H. Chen, Z. B. Li, S. Z. Deng, and N. S. Xu, *Quantum-mechanical investigation of field-emission mechanism of a micrometer-long single-walled carbon nanotube*, Phys. Rev. Lett. **92**, 106803 (2004).

[31]   A. Maiti, J. Andzelm, N. Tanpipat, and P. Von Allmen, *Effect of adsorbates on field emission from carbon nanotubes*, Phys. Rev. Lett. **87**, 155502 (2001).

[32] V. I. Kleshch, V. Porshyn, A. S. Orekhov, A. S. Orekhov, D. Lützenkirchen-Hecht, and A. N. Obraztsov, *Carbon single-electron point source controlled by Coulomb blockade*, Carbon **171**, 154 (2021).

[33] K. Nagaoka, H. Fujii, K. Matsuda, M. Komaki, Y. Murata, C. Oshima, and T. Sakurai, *Field emission spectroscopy from field-enhanced diffusion-growth nano-tips*, Appl. Surf. Sci. **182**, 12 (2001).

[34] S. Logothetidis, *Optical and electronic properties of amorphous carbon materials*, Diam. Relat. Mat. **12**, 141 (2003).

[35] S. T. Purcell, V. T. Binh, and R. Baptist, *Nanoprotrusion model for field emission from integrated microtips*, J. Vac. Sci. Technol. B **15**, 1666 (1997).

[36] M. Kozak, J. McNeur, K. J. Leedle, H. Deng, N. Schonenberger, A. Ruehl, I. Hartl, J. S. Harris, R. L. Byer, and P. Hommelhoff, *Optical gating and streaking of free electrons with sub-optical cycle precision*, Nat. Commun. **8**, 14342 (2017).

[37] K. E. Priebe, C. Rathje, S. V. Yalunin, T. Hohage, A. Feist, S. Schaer, and C. Ropers, *Attosecond electron pulse trains and quantum state reconstruction in ultrafast transmission electron microscopy*, Nat. Photonics **11**, 793 (2017).

[38] T. J. A. Wolf, D. M. Sanchez, J. Yang, R. M. Parrish, J. P. F. Nunes, M. Centurion, R. Coffee, J. P. Cryan, M. Guhr, K. Hegazy, A. Kirrander, R. K. Li, J. Ruddock, X. Shen, T. Vecchione, S. P. Weathersby, P. M. Weber, K. Wilkin, H. Yong, Q. Zheng, X. J. Wang, M. P. Minitti, and T. J. Martinez, *The photochemical ring-opening of 1,3-cyclohexadiene imaged by ultrafast electron diffraction*, Nat. Chem. **11**, 504 (2019).

Victor I. Kleshch[1,*], Vitali Porshyn[2], Pavel Serbun[2], Anton S. Orekhov[1,3,4], Rinat R. Ismagilov[1], Sergey A. Malykhin[1,5,6], Valentina A. Eremina[4,7], Elena D. Obraztsova[4,7], Dirk Lützenkirchen-Hecht[2], and Alexander N. Obraztsov[1,5]

[1]*Department of Physics, Lomonosov Moscow State University, Moscow 119991, Russia*

[2]*Physics Department, Faculty of Mathematics and Natural Sciences, University of Wuppertal, Wuppertal 42119, Germany*

[3]*NRC Kurchatov Institute, 123182 Moscow, Russia*

[4]*Moscow Institute of Physics and Technology, 141701 Dolgoprudny, Moscow Region, Russia*

[5]*Department of Physics and Mathematics, University of Eastern Finland, 80101 Joensuu, Finland*

[6]*Department of Solid State Physics, Lebedev Physical Institute, RAS, 119991 Moscow, Russia*

[7]*Prokhorov General Physics Institute, RAS, 119991 Moscow, Russia*

\* Corresponding author, e-mail: klesch@polly.phys.msu.ru

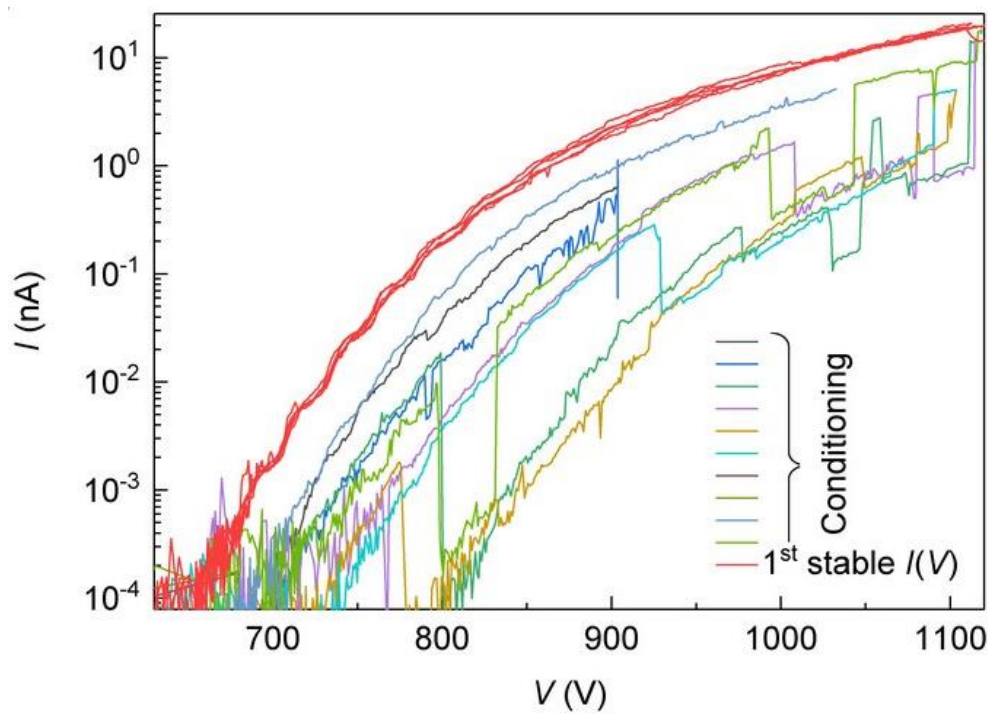

**Fig. S1.** Field emission (FE) current measurements performed in the scanning FE microscope (SFEM) for the SWCNT film during successive voltage cycles, demonstrating the conditioning process and the appearance of staircase-like current-voltage curve (shown in red) after several voltage ramps.

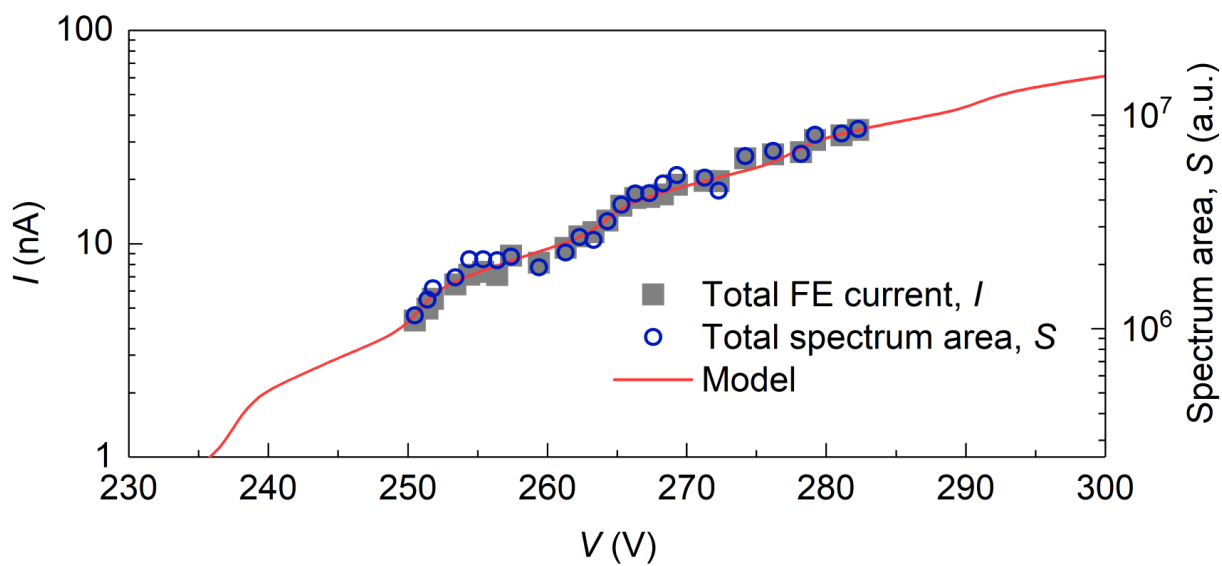

**Fig. S2.** Experimental (gray squares) and simulated (red line) current-voltage characteristics presented in Fig. 4a of the main text. Blue circles show the dependence of the total area under the spectra (before normalization) shown in Fig. 3a.

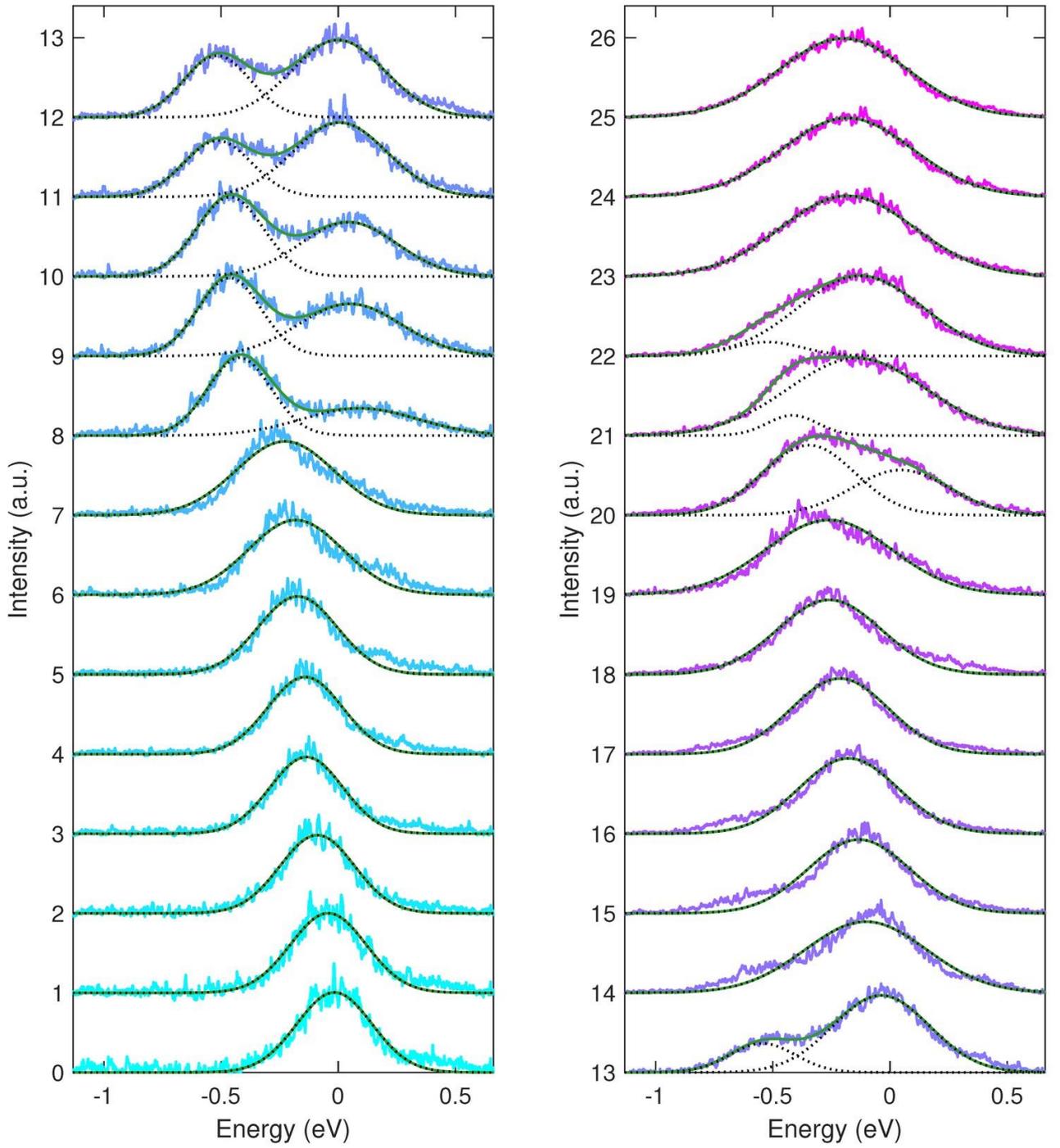

**Fig. S3.** Fits of the energy spectra presented in Fig. 3a of the main text. The dashed lines show fits of individual peaks using Gaussian functions $J(\varepsilon) = C_1 \exp\left(-(\varepsilon - \varepsilon_{peak})^2/C_2\right)$, where $\varepsilon_{\text{peak}}$, $C_1$, $C_2$ are the fitting coefficients. The resulting fits are shown by the green solid lines.